\documentclass[%
reprint,
superscriptaddress,
amsmath,amssymb,
aps,
prl,
floatfix,
longbibliography
]{revtex4-2}
\usepackage{graphicx} 
\usepackage{amsmath}
\usepackage{xcolor}
\usepackage[colorlinks=false]{hyperref}
\usepackage[margin=0.5in]{geometry} 
\usepackage{caption} 
\usepackage[official]{eurosym}
\usepackage{comment}
\usepackage{booktabs}
\usepackage{cases}
\usepackage{arydshln}
\usepackage{mathtools}

\definecolor{measuredColor}{rgb}{0.094, 0.631, 0.466}
\definecolor{estimatedColor}{rgb}{0.906, 0.623, 0.149}
\definecolor{calibratedColor}{rgb}{0.043, 0.454, 0.71}
\definecolor{fixedColor}{rgb}{0.835, 0.384, 0.16}

\begin{document}

\title{Farm Size Matters: A Spatially Explicit Ecological-Economic Framework\\ \quad \,for Biodiversity and Pest Management \, \quad}

\author{Elia Moretti}
\affiliation{Chair of Econophysics and Complex Systems, \'Ecole polytechnique, 91128 Palaiseau Cedex, France}
\affiliation{LadHyX, CNRS, École polytechnique, Institut Polytechnique de Paris, 91120 Palaiseau, France}

\author{Michel Loreau}
\affiliation{Station d'écologie théorique et expérimentale, CNRS, 09200 Moulis, France }
\affiliation{Institute of Ecology, College of Urban and Environmental Sciences, Peking University, Beijing 100871, China}

\author{Michael Benzaquen}
\affiliation{Chair of Econophysics and Complex Systems, \'Ecole polytechnique, 91128 Palaiseau Cedex, France}
\affiliation{LadHyX, CNRS, École polytechnique, Institut Polytechnique de Paris, 91120 Palaiseau, France}
\affiliation{Capital Fund Management, 23 rue de l'Université, 75007 Paris, France}

\date{January 2025}

    \begin{abstract}
    
          The intensification of European agriculture, characterized by increasing farm sizes, landscape simplification and reliance on synthetic pesticides, remains a key driver of biodiversity decline. While many studies have investigated this phenomenon, they often focus on isolated elements, resulting in a lack of holistic understanding and leaving policymakers and farmers with unclear priorities. This study addresses this gap by developing a spatially explicit ecological economic model designed to dissect the complex interplay between landscape structure and pesticide application, and their combined effects on natural enemy populations and farmers' economic returns. In particular, the model investigates how these relationships are modulated by farm size – a crucial aspect frequently overlooked in prior research. By calibrating on the European agricultural sector, we explore the ecological and economic consequences of various policy scenarios. We show that the effectiveness of ecological restoration strategies is strongly contingent upon farm size. Small to medium-sized farms can experience economic benefits from reduced pesticide use when coupled with hedgerow restoration, owing to enhanced natural pest control. In contrast, large farms encounter challenges in achieving comparable economic gains due to inherent landscape characteristics. This highlights the need to account for farm size in agri-environmental policies in order to promote biodiversity conservation and agricultural sustainability.
    \end{abstract}

\hfill \break

\maketitle

\section*{Introduction}

The decline of biodiversity in farmland ecosystems represents a significant environmental challenge, evidenced by numerous studies~\cite{ipbes2019global, StateofNature2020} and impacting various taxa~\cite{bird_decline, butterflies_decline}. Among these, insect populations have been particularly affected~\cite{insect_decline1, insect_decline2}, raising concerns due to their crucial roles in pollination~\cite{pollinator_decline}, pest control~\cite{pest_damage_climatechange}, and as a food source within the food web~\cite{insect_as_ecosystem}. Scientific literature points to two primary drivers of this decline~\cite{insect_decline1}: the simplification of agricultural landscapes through land consolidation and the widespread use of synthetic pesticides.

Driven by the need to meet increasing food demands and navigate intense international competition~\cite{agri_demand_intensification}, European agriculture has witnessed a consistent trend towards larger farm sizes over recent decades~\cite{land_consolidation,land_consolidation_2}. This shift, aimed at enhancing productivity and increasing income through the adoption of mechanized labor, has led to a simplification of the agricultural landscape. The expansion of field sizes, removal of linear landscape features such as hedgerows, and reduction of permanent grasslands have diminished the availability of diverse habitats essential for a rich array of species~\cite{field_biodiversity}. Simultaneously, the extensive application of synthetic pesticides, intended to maximize yields by controlling crop pests, has had significant side-effects, directly impacting beneficial insect populations and raising concerns about human health and environmental contamination~\cite{pesticide_biodiversity, pesticide_biodiversity_2}.

The unintended consequences of these practices are now becoming increasingly apparent. The reduction in the number of natural enemies of crop pests, coupled with the emergence of pesticide-resistant pest populations, has begun to contribute to a worrying trend of stagnating yields in some agricultural systems~\cite{stagnating_yield, stagnating_yield_2, stagnating_yield_3}. Furthermore, in the face of ongoing climate change, many scientific projections suggest that the vulnerability of these simplified and chemically reliant agricultural systems will likely be exacerbated, potentially leading to further ecological and economic instability if current farming practices persist~\cite{pest_damage_climatechange}.

In response to these challenges, policymakers and the scientific community have initiated efforts to promote and implement various policies aimed at mitigating these detrimental trends~\cite{conservation_poliocy_EU}.  However, the effectiveness of these policies has been limited thus far, as insect decline continues unabated, and the implementation of mitigation strategies often encounters strong opposition from the agricultural sector~\cite{ecological_intensification_problem}.

\begin{figure*}[t!]
\centering
\includegraphics[width=0.8\textwidth]{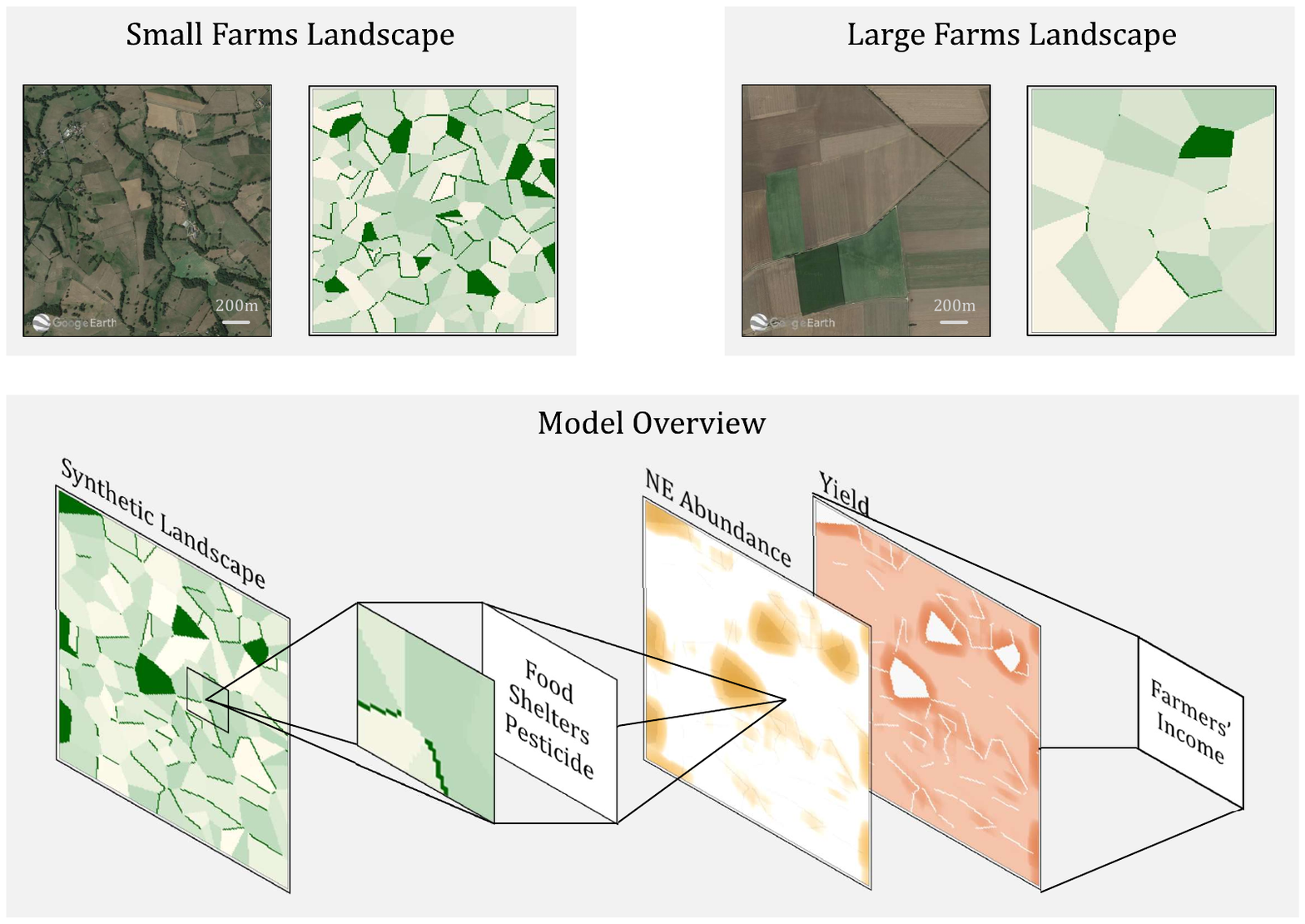}
\caption{ Conceptual scheme illustrating the landscape representation and the factors influencing natural enemy (NE) abundance, yield and farmers' income. The top panel shows how our synthetic landscape mimics the characteristics of real agricultural landscapes. Specifically, the left satellite imagery represents a typical small-farm landscape from the South of France near Limoges, while the right imagery depicts a characteristic large-farm landscape from Central Germany near Leipzig. The bottom panel visualizes the process by which we derive natural enemy abundance and farmers' income from a given landscape. As highlighted in the expanded view, the natural enemy (NE) abundance in each individual cell of the landscape grid is determined by considering the conditions within the typical movement range of such arthropods. Specifically, for each cell, we assess the spatial distribution of available food resources, the presence of shelters, and the intensity of pesticide application within that defined movement radius. The combined influence of these factors determines the resulting spatial pattern of NE abundance, visualized in the ``NE Abundance" layer, where darker orange signifies higher concentrations. The NE abundance, in conjunction with pesticide use, subsequently influences crop yield, which ultimately contributes to the computation of farmers' income.}
\label{fig:scheme}
\end{figure*}

While a fundamental paradigm shift is required to halt the general decline of biodiversity~\cite{paradigm_shift_book, paradigm_shift_montoya}, we also contend that three context-specific factors have hampered the success of current policies promoting sustainable pest management. Firstly, there is no clear consensus on  how management practices, especially pesticide use, and landscape composition jointly influence the development of natural enemy arthropods. Often, scientific analyses focus on isolated interventions~\cite{landscape_pesticide_review, landscape_pesticide_analysis}, failing to recognize the complex interplay between landscape and management factors. This lack of an integrative perspective leaves policymakers and farmers uncertain about priorities. Second, although numerous ecological studies propose seemingly viable paths towards ecological intensification – where natural processes substitute for synthetic inputs without reducing farmers' income – they frequently neglect the crucial economic realities, particularly at the farm level where policies are ultimately put into practice~\cite{ecological_intensification_problem}. Consequently, these policies often face resistance from various economic actors, resulting in diluted actions that have proven ineffective in reversing biodiversity decline. Third, a pivotal factor often neglected in both ecological and policy considerations is the significant heterogeneity of farm size within the agricultural sector~\cite{field_biodiversity}. Indeed, the agricultural sector exhibits a stark dichotomy~\cite{farm_size_distribution}: large farms, characterized by high profit margins but low ecological value, coexist with small farms, which provide significant ecological reservoirs but struggle with low profitability. As a consequence, a single strategy cannot be effective in both situations simultaneously.

To address these limitations, this paper introduces a theoretical framework designed to analyze the simultaneous effects of multiple policy instruments on both the ecological and economic outcomes within agricultural systems, with a specific focus on natural enemy abundance and the provision of natural pest protection. A crucial element of this framework is its spatially explicit description, aligning with a growing body of literature that recognizes the importance of landscape heterogeneity in supporting biodiversity and ecosystem services in agricultural systems, see e.g.~\cite{landscape_NE,landscape_NE_2, habitat_fragmentation_montoya, landscape_connectivity, habitat_percolation_paz}. In contrast to many existing models, which either employ a spatially explicit description while overlooking farm management practices~\cite{pest_potential} or utilize Lotka-Volterra models for pest control while neglecting landscape configuration~\cite{pesticide_lotka_volterra}, our model explicitly integrates both the spatial arrangement of diverse land cover types and the influence of management practices. Furthermore, compared with the limited number of studies that integrate both of these elements~\cite{ecological_economic_royal, pesticide_landscape_book}, we offer a more mechanistic understanding of these relationships by adopting a parsimonious approach that strategically focuses on the key factors influencing natural enemy abundance.

Furthermore, our framework is coupled with an economic description which makes it possible to simultaneously analyze the impacts of policies on both ecological and economic outcomes, such as farmers' income. A crucial element of this integrated framework is the explicit differentiation between farm sizes. Recognizing that farms of varying scales exhibit distinct landscape characteristics~\cite{field_biodiversity}, management practices~\cite{farm_size_sustainability}, and economic priorities~\cite{farm_size_economics} is essential for developing effective and equitable policy recommendations. With this work, we aim to answer the following key questions: Which combinations of policy instruments are most effective in halting the decline of arthropod populations across different farm sizes? And under what spatial and management scenarios can these policies be implemented in a way that leads to positive or at least minimally disruptive economic outcomes?

\section*{Model and methods}

\subsection*{Model overview}

We formulate a spatially explicit model that describes the abundance of predatory arthropods and its relationship with farmers' income (Fig.~\ref{fig:scheme}).
The model operates on a discrete spatial representation of the farmland landscape, conceptualized as a square periodic lattice. Each cell  within this lattice represents a specific land-cover type, which can be either cropland, hedgerow, or permanent grassland. The spatial arrangement and proportion of these land-cover elements are based on farm size, reflecting the observed tendency for farms of different scales to exhibit distinct landscape characteristics. Indeed, larger farms often feature larger fields and a lower proportion of semi-natural habitat (SNH) than do smaller farms.

The dynamics of natural enemy abundance within this spatially defined landscape are modeled using a seasonal logistic growth framework. This ecological component explicitly incorporates the impact of both pesticide application and the landscape on the availability of overwintering sites, food resources, and mortality rates induced by pesticide exposure.

The abundance of predatory arthropods is then used to derive farmers' yield in conjunction with pesticide use. This establishes a direct pathway through which ecological factors influence the economic performance of farming operations. All other economic data required for the calculation of farmers' income, such as crop prices, operational costs, subsidies, and labor input, are incorporated as exogenous variables derived from the European Farm Accountancy Data Network (FADN), providing an empirical basis for the economic simulations.

We then use the model to study how the spatial configuration of the agricultural landscape, in combination with different pesticide use regimes, impacts both the abundance of beneficial predatory arthropods and the resulting economic outcomes for farmers. In what follows, we provide detailed descriptions of the individual components of the model, including the process of landscape construction, the ecological model of natural enemy dynamics, and the economic model used to assess farmers' income.

\subsection*{Landscape construction}

The landscape consists of a two-dimensional lattice ($200 \times 200$ cells), in which each cell represents an area of $10\mathrm{m} \times 10\mathrm{m}$ and can be in one of two states: cropland or SNH. To determine the allocation of each cell, we begin by assuming that the entire landscape is covered with cropland and then assign each cell to a unique field using a Voronoi tessellation, a method widely employed for creating semi-realistic field distributions~\cite{tasselization_field}. For clarity, let us stress that in the following discussion, ``cells" always refer to individual ``pixels", whereas  the larger Voronoi cells that comprise multiple pixels will be referred to as ``fields". The pattern generation is controlled by a parameter describing the desired mean field size (see Supplementary Material for more details).

\begin{figure*}[t!]
\centering
\includegraphics[width=\textwidth]{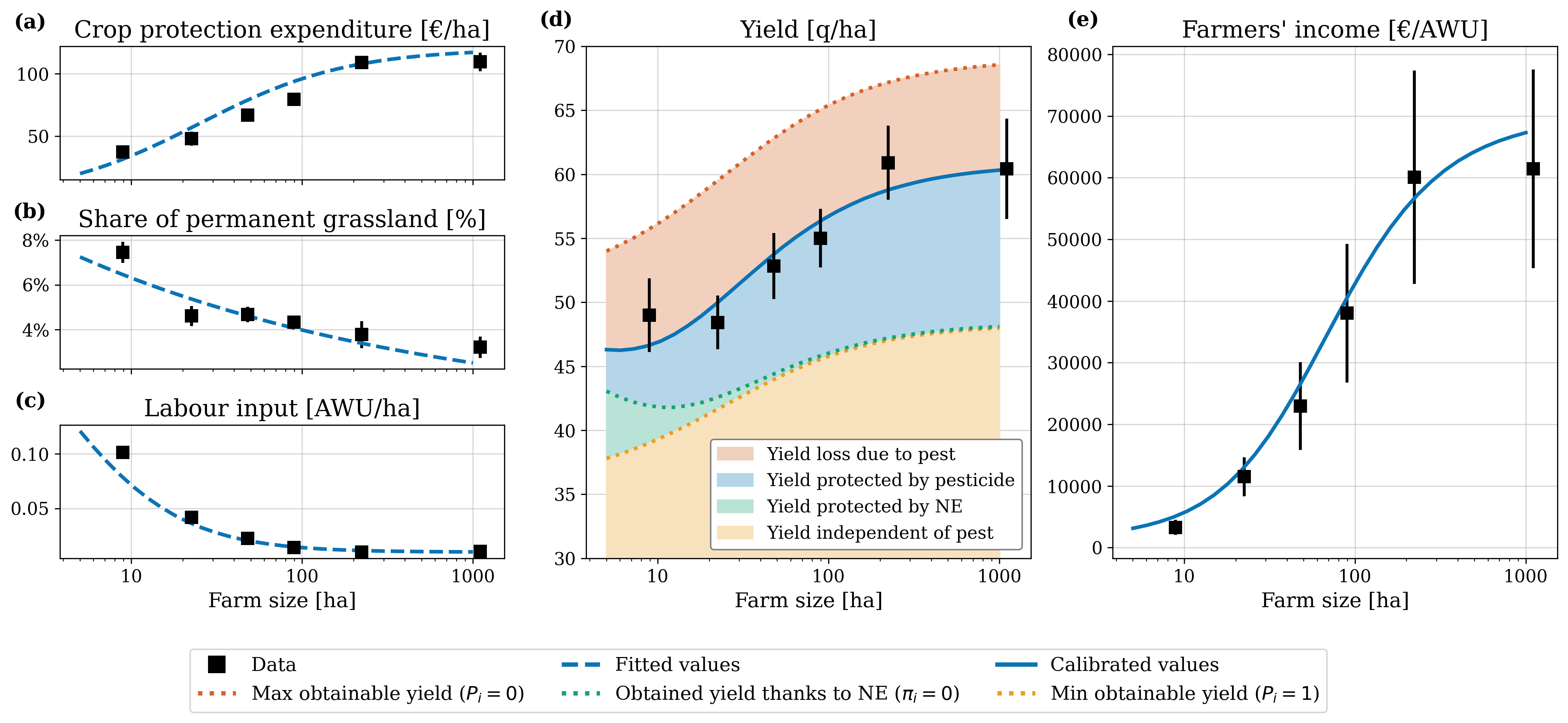}
\caption{This figure illustrates the calibration of our model's parameters using aggregated agricultural data from the European Farm Accountancy Data Network (FADN) for the EU-27 region, averaged over the years 2018-2022~\cite{FADN}. Black squares represent the weighted average of the observed data across this period, with error bars indicating the standard deviation of year-on-year variation. These data highlight consistent patterns across farm size for different year and regions, including increased reliance on synthetic inputs, a reduced share of grassland, and higher income for large farms relative to small ones. The parameterization process differed based on the variable's dependence on ecological dynamics. For variables independent of these dynamics (examples shown in the left panel), we employed a fitting procedure. Conversely, for variables directly influenced by ecological dynamics (examples in the central and right panels), we utilized a calibration process that minimized a quadratic loss function (further details on the calibration and additional data not displayed here are available in the Supplementary Materials). The central panel, depicting yield, provides a decomposed view: the maximum obtainable yield ($P_i=0$ in Eq.~\ref{eq:yield}), the yield protected by natural enemies ($\pi_i=0$ in Eq.~\ref{eq:pest_damage}), and the minimum obtainable yield ($P_i=1$ in Eq.~\ref{eq:yield}). This disaggregation highlights that smaller farms tend to benefit more significantly from natural pest protection, while larger farms appear to rely predominantly on synthetic inputs for pest protection.}
\label{fig:calibration}
\end{figure*}

We then assign SNH by converting some cells into either permanent grassland or hedgerows. A key constraint in the placement of hedgerows is that they are exclusively located along the margins of the previously defined crop fields, reflecting their typical occurrence in real agricultural landscapes.  The allocation of grassland and hedgerows is managed by an iterative algorithm. In each iteration, the entire area or a portion of a field (or field margin) is converted into the designated SNH type. This process continues until the total area converted to grassland and hedgerows reaches $gA$ and $hA$, respectively, where $A$ is the total landscape area, $g$ is the grassland share, and $h$ is the hedgerows share. Although $g$ and $h$ are adjustable parameters, when analyzing landscapes with a specific farm size, we can define them as functions of that size, a relationship strongly supported by existing research and data (see Fig.~\ref{fig:calibration} and Supplementary Material for more details).


\subsection*{Natural enemies' abundance}

Our model provides a simplified description of flying generalist natural enemies (NE) (i.e., ladybirds) feeding on herbivorous insect pests (i.e., aphids) in annual crop systems (i.e., winter wheat).
While pest dynamics are only implicit, we explicitly model the NE lifecycle, distinguishing between summer and winter. 

During the summer, NE disperse from their overwintering locations and colonize cereal fields in search for food. The abundance of NE in each grid cell is modeled using logistic growth, incorporating food availability and pesticide-related mortality in the patch and its surroundings:
\begin{equation}
    \frac{dN_i}{dt} = r N_i \left(1 - \frac{N_i}{k_i} \right) - N_i m_i
\end{equation}
where $r$ is the reproduction rate, $k_i$ is the carrying capacity, and $m_i$ is an extra-mortality rate due to pesticide use. One unit of time $t$ corresponds to one day during the growing season, which we assume extends for six months. 

We define the carrying capacity ($k_i$) based on the availability of food resources, specifically herbivorous insects, which is influenced by both pesticide use and land cover type. This relationship is expressed by the following equation:

\begin{equation}
\label{eq:carrying_capacity}
    k_i = \sum_j \gamma_{ij} \kappa_j \quad \text{where} \quad
    \kappa_j =
    \begin{dcases*}
    \kappa_{\mathrm{SN}} & \text{if  $\tau_{j} = \mathrm{SN}$,} \\ 
    \kappa_{\mathrm{C}} \cdot \mathcal{F}\left[\frac{\pi_j}{\pi_\mathrm{ref}}\right] & \text{if  $\tau_{j} = \mathrm{C}$.}
    \end{dcases*}
\end{equation}

Here, $\kappa_{\mathrm{SN}}$ and $\kappa_{\mathrm{C}}$ represent the maximum predatory arthropod density (number per cell) in the absence of pesticide application for SNH and cropland (C), respectively. The function $\mathcal{F}$ is defined as $\mathcal{F}\left[\frac{x}{y}\right] \equiv \frac{1}{1+\frac{x}{y}}$, and $\pi_j$ represents the amount of pesticide applied to patch $j$. In particular, we assume that pesticide application occurs uniformly and exclusively on cropland patches, thereby neglecting potential drift effects into SNH. Finally, $\gamma$ is a distance-decay function which represents the movement range of NE over a year. Specifically, we employed a Heaviside step function as a computationally efficient approximation of an exponential decay, which is consistent with both theoretical and empirical estimations~\cite{pest_potential, distance_decay_function, abm_pest}. We set:

\begin{equation}
    \gamma_{i, j} = \begin{dcases*}
        1 \quad \text{  if  } \quad d_{i,j}\leq D \\
        0 \quad \text{  if  } \quad  d_{i,j}>D
    \end{dcases*}
\end{equation}
where $d_{i,j}$ represents the distance between cells $i$ and $j$, and $D$ is the parameter defining the typical annual movement range of  natural enemies.

The additional mortality experienced by arthropods in patch $i$ due to pesticide exposure is a function of the total amount of pesticide applied within its movement kernel, $\gamma$. Specifically:
\begin{equation}
\label{eq:extra_pest}
    m_i = \mu_\pi \cdot \mathcal{F}\left[\frac{\pi_\mathrm{ref}}{\sum_j \gamma_{ij} \pi_j}\right]
\end{equation}
where $\mu_\pi = (1 - q)r$ is the induced mortality rate, and the parameter $q$  modulates the selectivity of the pesticide application in affecting NE or pest population only. A higher $q$ indicates greater selectivity, meaning pesticides have a stronger negative impact on pests relative to their natural predators, because the effect on the carrying capacity in Eq.~\ref{eq:carrying_capacity} is more pronounced than the increase in the extra-mortality rate in Eq.~\ref{eq:extra_pest}. Conversely, a lower $q$ suggests that the pesticide is less selective, affecting both natural enemies and herbivorous arthropods.

As the season progresses and temperatures begin to drop, NE start to prepare for hibernation, seeking shelters to overwinter. The presence of SNH determines the success of this process~\cite{SNH_overwinter}. If SNH are not large enough, a portion of the population will die, as implicitly expressed by the equation:
\begin{equation}
    N_i(\text{end of winter}) = (1 - s_i) N_i(\text{end of summer})
\end{equation}
    
where $s_i$ is the survival rate that depends on the presence of semi-natural habitats:
\begin{equation}
    s_i = \mathcal{F}\left[\frac{\mathrm{sn}_\mathrm{ref}}{\mathrm{sn}_i}\right]  \quad \text{where} \quad
    \mathrm{sn}_i = \sum_j \gamma_{ij} \cdot \delta( \tau_j = \mathrm{SN})
\end{equation}

\subsection*{Farmers' economic performance}

To compute farmers' economic performance, we do not consider within-year dynamics but focus on the amount of crop biomass produced annually ($y_i$), i.e., annual crop yield, which then determines farmers' income across different farm sizes.
In our model, $y_i$ is not represented by a differential equation, as crops are harvested each year and their dynamics do not depend on the previous state. But $y_i$ depends on the amount of pesticide used and the abundance of NE, as represented by the following equation:
\begin{equation}
\label{eq:yield}
    y_i=\left(y_0+y_1\mathcal{F}\left[\frac{L}{L_{\mathrm{ref},y}}\right]\right)\left(1-\eta P_i\right)
\end{equation}

In this equation first contribution that sets the maximum potential yield while the second describes the yield loss due to pest exposure. In the first term, $y_0$ represents the baseline yield achievable in the absence of intensification practices (such as in organic farming~\cite{organic_yield}) and $y_1$ represents the increase in yield resulting from the use of fertilizer, which increases with farm size $L$~\cite{farm_size_fertilizer}. In the second term, $\eta$ represents the maximum loss fraction due to pest damage, and $P_i$ is the pest damage. 

To determine pest damage, we consider the impact of pesticide and NE according to:
\begin{equation}
\label{eq:pest_damage}
    P_i=1-\mathcal{F}\left[\frac{\pi_{\mathrm{ref}}}{\pi_i}\right]-\frac{N_i}{\kappa_\mathrm{C}}=\mathcal{F}\left[\frac{\pi_i}{\pi_{\mathrm{ref}}}\right]-\frac{N_i}{\kappa_\mathrm{C}}
\end{equation}
    
where $N_i$ represents NE abundance at the end of the summer and $\pi_i$ depends on farm size according to $\pi_i = \pi_0 \mathcal{F}\left[\frac{L_{\mathrm{ref, \pi}}}{L}\right]$. Here, we assume that an increase in either the presence of natural enemies or pesticide application leads to higher yields. 

Farmers' profit is then given by:
\begin{equation}
    \mathcal{P} = p L_{\mathrm{prod}} \bar{y}  - \mathcal{C} + L\mathcal{S}
\end{equation}
where $p$ is crop price, $\mathcal{C}$ operational costs, $\mathcal{S}$ subsidies per hectare, and $L_{\mathrm{prod}} =L-gL-hL$ is the area dedicated to crops. Operational costs increase proportionally to the total land area, including land dedicated to ecological purposes, which requires maintenance:
\begin{equation}
    \mathcal{C}=\left(\mathcal{C}_0+\pi\right)L_{\mathrm{prod}}+\mathcal{C}_\mathrm{g}Lg+\mathcal{C}_\mathrm{h}Lh 
\end{equation}
where  $\mathcal{C}_0$ are operational costs per hectare including fuel, fertilizer, and depreciation of machinery; and $\mathcal{C}_\mathrm{g}$ and $\mathcal{C}_\mathrm{h}$ are maintenance costs per hectare for hedgerows and grassland, respectively.

A good proxy for farmers’ income, consistent with European institutional practices, can then be computed as the ratio between total profits $\mathcal{P}$ and labor input $\mathcal{L}$~\cite{Farm_income_def}:
\begin{equation}
    \mathcal{I} = \frac{\mathcal{P}}{\mathcal{L}}
\end{equation}
    
where $\mathcal{L}$ decreases with farm size $L$, reflecting improved management efficiency, and field size $S$  due to the use of larger machinery:
\begin{equation}
        \mathcal{L} = L_{\mathrm{prod}} \left(\ell_0 + \ell_1 \mathcal{F}\left[\frac{L_{\mathrm{prod}}}{L_\mathrm{ref, \mathcal{L}}}\right] \mathcal{F}\left[\frac{S}{S_\mathrm{ref}}\right] \right)
\end{equation}

\subsection*{Parameter specification}

The model parameters were specified through a combination of data-driven estimation, calibration against observed agricultural performance, and fixed values set to explore different scenarios.

Ecological parameters, such as natural enemy reproduction ($r$), carrying capacities ($\kappa_{\mathrm{SN}}, \kappa_{\mathrm{C}}$), and dispersal distance ($D$), were primarily estimated using empirical information and commonly-assigned values found in the literature~\cite{ecological_economic_royal, habitat_fragmentation_montoya, carrying_capacity_parameter, carrying_capacity_parameter_2}. However, for some of the parameters for which the literature does not provide a single value, such as pesticide application selectivity ($q$) or winter survival ($\mathrm{sn}_\mathrm{ref}$), we adopted reasonable starting values and subsequently performed sensitivity analyses (detailed in the Supplementary Materials) to evaluate their influence on the results, given their inherent context specificity and complexity.

Economic parameters were determined using 2018-2022 EU-27 average data from the European Farm Accountancy Data Network (FADN)~\cite{FADN}. This allowed us to capture trends across farm sizes and countries. Parameters independent of ecological dynamics (as crop prices $p$, operational costs $\mathcal{C}_0$, subsidies $\mathcal{S}$, and labor input $\mathcal{L}$), were fitted directly to the FADN data. Parameters affecting crop yield ($y_0, y_1, L_{\mathrm{ref},y}$, $\pi_{\mathrm{ref}}$) were calibrated to reproduce the observed trends in yield, economic outputs, and farmers' income (with some shown in Fig.~\ref{fig:calibration} and the complete list of economic outputs available in the Supplementary Materials).

\subsection*{Computational implementation}

We conducted extensive numerical simulations to explore various policy scenarios. For each simulation, the abundance of NE was computed over an initial equilibration period of 10 years to ensure the system reached a stable stationary state. NE abundance reported below corresponds to NE abundance at the end of the summer season, after the equilibration period. Once NE abundance stabilized, we used it to derive all relevant economic variables. 
The results presented below are averages across 100 simulations. Relative standard errors were omitted from figures due to their negligible magnitude (smaller than the line width). It is important to note that our analysis did not incorporate stochasticity in either the ecological or economic components of the model; therefore, the minor variations observed between simulations arose solely from the inherent variability among the randomly generated landscape configurations.

\begin{figure*}[t!]
\centering
\includegraphics[width=\textwidth]{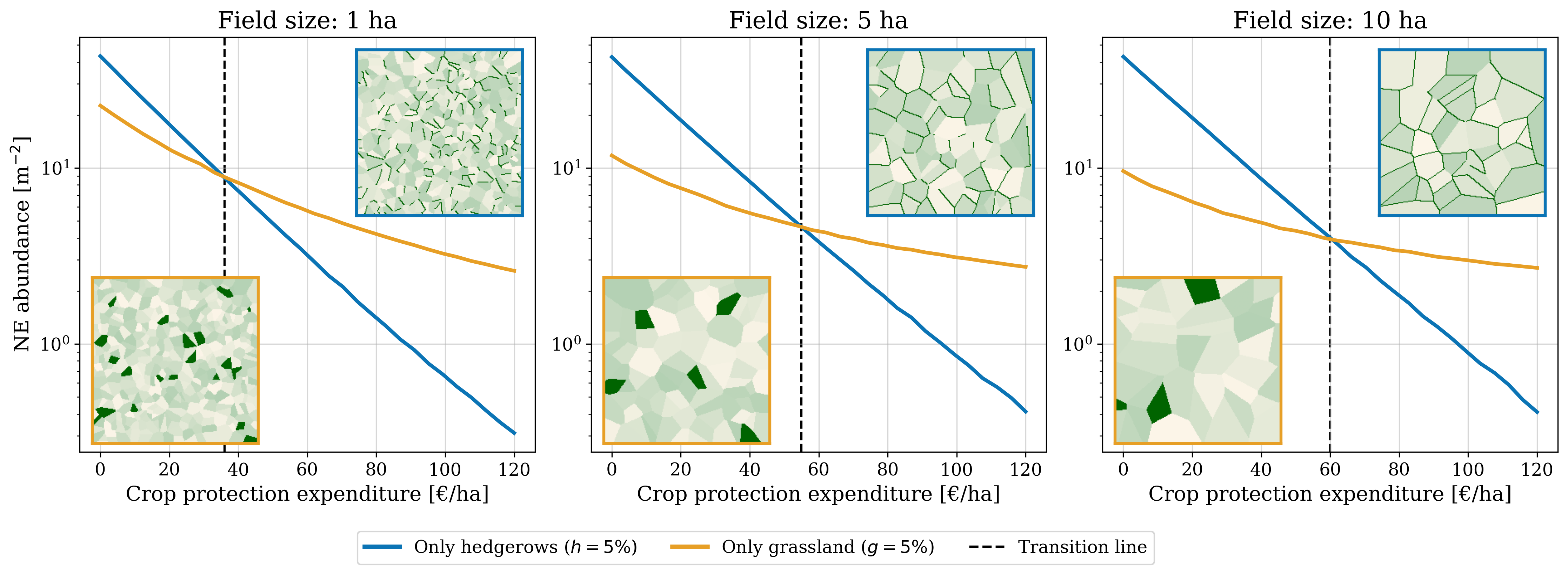}
\caption{Influence of landscape structure and crop protection expenditure on natural enemy (NE) abundance across different field sizes. This figure illustrates the simulated impact of crop protection expenditure (\euro{}/ha), serving as a proxy for pesticide use intensity, on natural enemy abundance (NE abundance [m$^{-2}$]) across three different field sizes: 1 hectare (ha), 5 ha, and 10 ha. Within each panel, two distinct landscape scenarios are compared: a landscape with 5\% coverage of semi-natural habitat (SNH) composed solely of hedgerows (blue line), and a landscape with 5\% coverage of SNH composed solely of permanent grassland patches (orange line). Representative spatial configurations of these two SNH types within each field size are depicted as insets within each panel. The figure demonstrate a general trend of decreasing NE abundance with increasing crop protection expenditure across all field sizes and SNH compositions, highlighting the negative impact of pesticide use. However, the relative effectiveness of different SNH compositions in sustaining NE abundance varies depending on both the intensity of crop protection and the average field size. At low levels of crop protection expenditure, landscapes with a higher proportion of hedgerows generally support greater NE abundance. Conversely, at higher levels of crop protection expenditure, landscapes with a higher proportion of permanent grassland tend to maintain relatively higher NE abundance. Interestingly, the transition point and the overall impact of landscape structure on NE abundance are modulated by field size. In landscapes with larger fields, the advantage of hedgerows is more significant and  grassland-dominated SNH composition only becomes favorable at higher levels of pesticide use.}
\label{fig:field}
\end{figure*}

\section*{Results}

\subsection*{Landscape structure, pesticide use, and natural enemy abundance}

We first examine the relationship between natural enemy (NE) abundance, landscape structure, and crop protection expenditure, which is an indicator of pesticide use intensity. As expected, increased pesticide use consistently leads to a reduction in NE populations across all landscape scenarios and field sizes examined (Fig.~\ref{fig:field}). This underscores the significant negative impact of pesticides on the decline of predatory arthropod populations~\cite{pesticide_biodiversity}.

However, the relative effectiveness of different SNH spatial arrangements in supporting NE populations is contingent upon the level of crop protection expenditure. Under conditions of low crop protection expenditure, landscapes with a more fragmented distribution of SNH, characterized by a greater prevalence of hedgerow-like features, generally exhibit higher NE abundance (blue lines in Fig.~\ref{fig:field}). This suggests that the increased edge habitats and connectivity provided by fragmented SNH enhance the availability of shelters, which are particularly important for sustaining large NE populations during overwintering. Conversely, when crop protection expenditure is high, landscapes with a more condensed SNH configuration, containing larger patches of permanent grassland, tend to support a relatively higher NE abundance (yellow lines in Fig.~\ref{fig:field}). This is because permanent grassland areas function as refuge zones, providing safer habitats where NE can avoid or minimize exposure to pesticides applied in adjacent croplands. 

Interestingly, the relative benefits of fragmented versus condensed SNH appear to shift depending on field size. The vertical dashed black line in Fig.~\ref{fig:field}, marking the pesticide expenditure level at which grassland become more beneficial than hedgerows for NE, shifts towards higher pesticide use as field size increases. Thus, in landscapes with larger fields, fragmented SNH (hedgerows) maintain a relative advantage for NE abundance even under higher pesticide pressure. This can be attributed to the fact that when grassland patches become larger than the typical movement range of NE, their benefits for NE become limited. Indeed, as the foraging activities of most organisms are localized, NE arthropods tend to thrive in heterogeneous landscapes that integrate both SNH and cropland at the scale of their typical movement range~\cite{habitat_fragmentation_montoya}. Furthermore, a more fragmented landscape facilitates the movement of NE from SNH into nearby crop areas, potentially boosting yield and farmers' income, as discussed in the following section.

\begin{figure*}[t!]
\centering
\includegraphics[width=\textwidth]{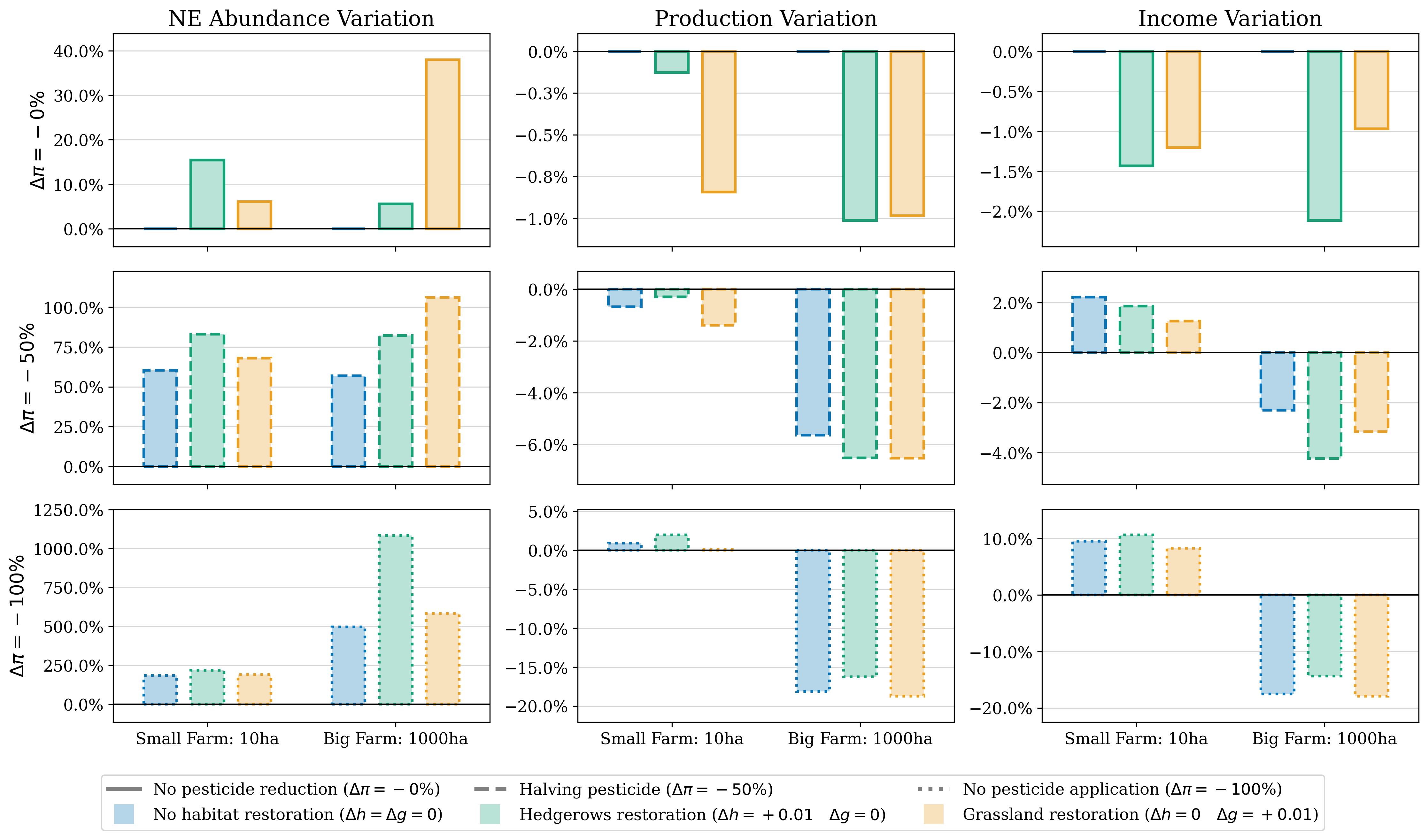}
\caption{Impact of pesticide reduction and semi-natural habitat (SNH) restoration policies on natural enemy (NE) abundance, production, and farmers' income across different farm sizes. The figure displays the percentage change in NE abundance (left column), production (center column), and farmers' income (right column) relative to a baseline scenario under three distinct pesticide reduction levels: no reduction ($\Delta \pi = -0\%$, top row), a 50\% reduction ($\Delta \pi = -50\%$, middle row), and a complete cessation of pesticide use ($\Delta \pi = -100\%$, bottom row). Within each row, results are shown for two farm sizes: a small farm (10 ha) and a large farm (1000 ha). For each pesticide reduction level and farm size, three different SNH restoration policies are compared: no change in SNH ($\Delta h = \Delta g = 0$, solid bars), a 1\% increase in hedgerow cover ($\Delta h = +0.01, \Delta g = 0$, dashed bars), and a 1\% increase in permanent grassland cover ($\Delta h = 0, \Delta g = +0.01$, dotted bars). Without any pesticide reduction, increasing hedgerows leads to a more substantial increase in NE abundance for small farms compared to increasing grasslands, while the opposite is true for large farms. By reducing pesticide reliance, NE abundance increases significantly for both farm sizes, leading to a positive income change for small farms while large farms still suffer from negative income variation. Moreover, hedgerow restoration appears to be the most effective strategy for minimizing income losses under a zero-pesticide regime across all farm sizes.} 
\label{fig:bar}
\end{figure*}

\begin{figure*}[t!]
\centering
\includegraphics[width=\textwidth]{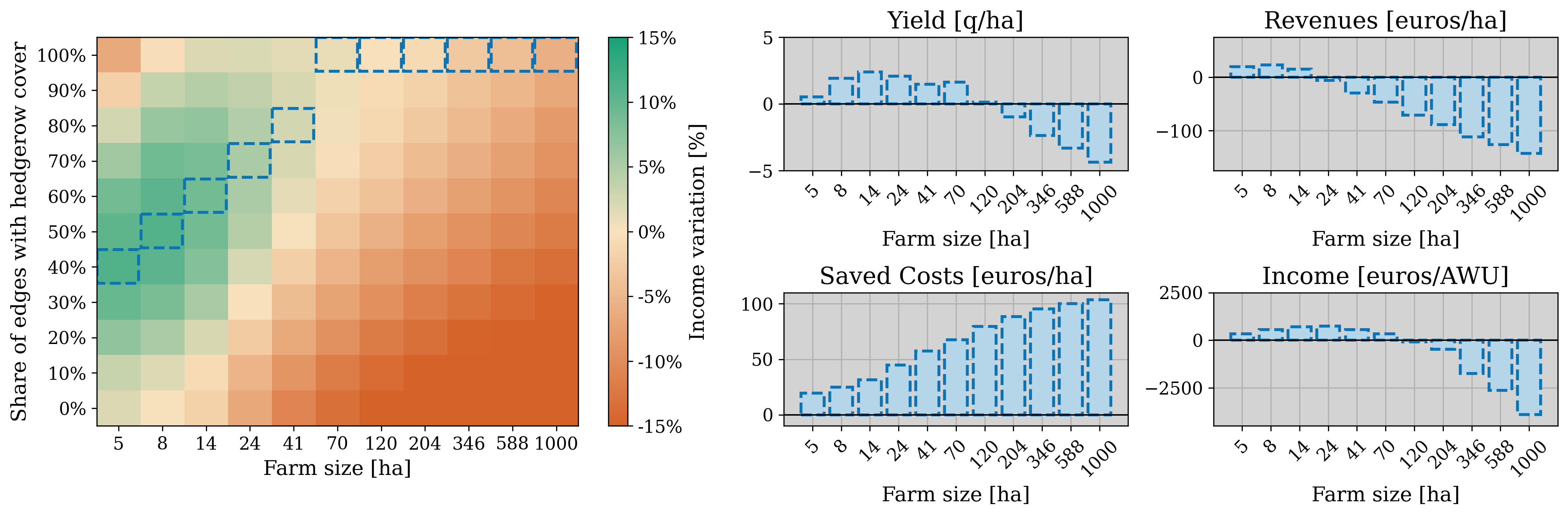}
\caption{Economic impact of transitioning to a zero-pesticide policy combined with varying hedgerow shares across a spectrum of farm sizes. The heatmap on the left illustrates the percentage variation in farmers' income relative to the baseline scenario (current pesticide use and semi-natural habitat levels) as a function of farm size (x-axis, logarithmic scale in hectares) and hedgerow share (y-axis, proportion of total field edges covered by hedgerows). The dashed blue lines highlight regions of maximum income variation, each for a single farm size. The panels on the right provide a more detailed breakdown of the changes in yield (q/ha), revenues (\euro{}/ha), saved pesticide costs (\euro{}/ha), and income (\euro{}/AWU) under the optimal hedgerow share for each farm size.  The figure demonstrates that ecological intensification  through a zero-pesticide policy and optimized hedgerow restoration can be economically beneficial for small to medium-sized farms. However, for large farms, while pesticide cost savings are substantial, they are insufficient to compensate for the potential revenue losses associated with slightly reduced yields in the absence of synthetic pesticides, highlighting the challenges of directly translating ecological benefits into economic gains at larger agricultural scales. }
\label{fig:phase_diagram}
\end{figure*}

\subsection*{Policy impacts on yield and income across farm sizes}

We now examine the potential effects of different policy interventions on NE abundance, crop yield, and farmers' income as a function of farm size, taking into account the structural differences that farm size imposes (large farms typically exhibit higher pesticide use and less SNH than do small farms). Fig~\ref{fig:bar} shows the percentage change in these three key variables under different policy scenarios: no pesticide reduction, a 50\% pesticide reduction, and a complete cessation of pesticide use, each combined with either no change in SNH or a 1\% increase in either hedgerow or grassland share.

\textit{No Pesticide Reduction:} For small farms, an increase in the proportion of hedgerows yields a greater enhancement in NE abundance than does an increase in the proportion of grasslands. The opposite trend is observed in large farms, as their lower SNH and their higher pesticide use make larger habitat patches more impactful. Notably, neither SNH increase policy leads to a net income increase for either farm size. In large farms, while increased NE abundance from grassland expansion boosts yield, this does not compensate for the reduced area of arable land, resulting in decreased total production and consequently, lower income. In small farms, although there is a small overall increase in NE abundance from hedgerow restoration leading to a higher yield (due to greater spillover effects than in grasslands), the added management costs associated with habitat restoration outweigh the production gains, leading to a net income decrease. 
    
 \textit{50\% Pesticide Reduction:} With a 50\% reduction in pesticide use, the patterns of NE abundance remain similar: hedgerow expansion benefits small farms more, and grassland expansion benefits large farms more. However, the income implications diverge. Small farms experience an increase in income under all policy scenarios. This is primarily attributed to the cost savings resulting from reduced pesticide expenditure. Among the strategies, maintaining the current SNH levels (no hedgerow or grassland restoration) proves most beneficial for small farmers' income, as the added management costs of habitat restoration offset the yield benefits. For large farms, none of the considered policies results in a net income increase. Indeed, the enhanced NE abundance does not fully compensate for the reduced pest control from synthetic inputs, resulting in lower yields that impact profits more significantly than the saved pesticide costs. 

\textit{No Pesticide Application:} Under a complete cessation of pesticide use, large farms exhibit a more substantial increase in NE abundance than do small farms, due to their lower initial NE populations. While small farms can achieve approximately a 10\% increase in income under this scenario, large farms still cannot achieve a positive income variation. Nevertheless, for both farm sizes, increasing hedgerows emerges as the most effective strategy to preserve both yield and income in the absence of pesticides, highlighting the crucial role of linear landscape features in supporting NE populations and their associated benefits when chemical pest control is eliminated.

\subsection*{Economic potential of natural pest protection without pesticides}

Building upon the previous finding that hedgerow restoration combined with zero pesticide use can enhance the economic performance of small farmers, we now investigate the potential for large farms to substitute synthetic inputs with natural processes without reducing their income. Specifically, we explore the optimal hedgerow restoration policy required to maximize economic performance under a zero-pesticide regime across the spectrum of farm sizes. 

Fig~\ref{fig:phase_diagram} illustrates the variation in income resulting from a shift from current practices (current pesticide use and SNH levels) to a zero-pesticide policy combined with varying proportions of hedgerows across different farm sizes. The most substantial positive income variations are evident for small to medium-sized farms, indicating that ecological restoration offers a more economically advantageous pathway than reliance on synthetic inputs at these scales. The combination of increased yield and cost savings (reduced pesticide expenditure minus management costs for increased hedgerows) contributes to an income increase ranging from 5\% for 50ha farms to 15\% for 5ha farms. 

For large farms, while a shift to a zero-pesticide policy with hedgerow restoration does not lead to a net benefit, the relative income decrease remains modest, at approximately -5\%. Such limited negative impact is primarily due to the significant cost savings that result from eliminating pesticide use, which partially offset the revenue loss due to lower yields. The latter can be attributed to the increase in NE abundance not fully compensating for the pest control efficacy previously provided by pesticides. This, in turn, is largely because large farms with larger fields inherently have a smaller proportion of edge habitat relative to their total area, limiting the potential for natural pest control to achieve the same level of effectiveness as in landscapes with smaller fields. 

\section*{Discussion}

The need for a more integrative understanding of pest management in agricultural landscapes stems from the increasing recognition of the ecological costs and sustainability challenges associated with conventional, input intensive farming practices~\cite{insect_decline1, ipbes2019global}. While the impact of pesticides on biodiversity and ecosystem services is well documented~\cite{pesticide_biodiversity, pesticide_biodiversity_2}, the complex spatial dynamics governing pest and natural enemy interactions, particularly in relation to the configuration of semi-natural habitats, are often overlooked or simplified in existing models. To address this critical gap, our study introduces a novel ecological perspective by employing a parsimonious yet spatially explicit model that links natural pest control to both pesticide application intensity and the spatial arrangement of different land cover types, particularly cropland and SNH. This allows us to provide a more realistic and context-dependent understanding of natural enemy abundance and its potential for pest control across varying farm sizes and policy scenarios.

The insights derived from this spatially explicit modeling approach highlights the importance of considering farm size as a critical factor in the design and implementation of agri-environmental policies aimed at promoting biodiversity and reducing reliance on synthetic inputs. Indeed, the contrasting responses observed in our analysis between small and large farms demonstrates that a one-size-fits-all approach is unlikely to be effective. While our model demonstrates the detrimental impact of conventional pesticide use on predatory arthropod populations, aligning with a substantial body of existing literature, it also shows that the effectiveness of ecological restoration strategies exhibits a strong dependence on farm size and the spatial configuration of SNH. This underscores the complex relationships between agricultural management practices, landscape structure, and both ecological and economic outcomes~\cite{landscape_pesticide_review, ecological_intensification_problem}.

For small to medium-sized farms, a transition towards reduced or eliminated pesticide use, coupled with targeted hedgerow restoration, presents a promising pathway for enhancing both farmland biodiversity and farmers' income. Hedgerows offer vital shelter for natural enemy arthropods, potentially increasing yields and reducing the need for costly synthetic inputs. However, despite this economic potential, such a transition is unlikely to occur spontaneously. Small farmers often operate with narrow profit margins~\cite{farm_size_economics}, which makes them particularly vulnerable to potential income losses in the early years of adopting new, ecologically based practices. Furthermore, their limited access to information networks and centralized policy discussions can breed reluctance to adopt unfamiliar strategies~\cite{small_farmer_barrier_sustainability}. This highlights a crucial role for policymakers in actively facilitating this shift~\cite{small_farms_for_sustainability}. Dedicated financial mechanisms are needed to buffer small farmers against the short-term economic risks associated with ecological transition, alongside proactive extension services and information campaigns that clearly demonstrate the long-term economic and ecological benefits of ecological restoration.

In contrast, the potential for large farms to achieve similar economic benefits from ecological restoration strategies appears more constrained. The inherent characteristics of large-scale agriculture, with a lower proportion of edge habitat, limit the effectiveness of natural pest control. While eliminating pesticide costs offers a significant economic advantage, the yield increases resulting from enhanced natural enemy abundance may not fully compensate for the reduced pest control efficacy previously provided by synthetic inputs. Consequently, our model suggests that large farms may experience a modest decrease in income under a complete transition to a zero pesticides policy and increased hedgerow cover. This result underscores the critical role of field size in mediating the benefits of ecological approaches and suggests that biodiversity support schemes should consider incentivizing the maintenance or creation of smaller field sizes, particularly in regions dominated by large-scale agriculture~\cite{field_biodiversity}. Moreover, this highlights the potential need to shift from purely individual-farm payments~\cite{CAP_history} towards more collaborative, landscape-level initiatives that encourage larger farms to support ecological enhancements on smaller neighboring farms~\cite{comunity_subsidies}, potentially fostering a more diverse and resilient agricultural mosaic while addressing concerns about land consolidation.

Our model, while providing valuable insights, exhibits certain limitations that suggest avenues for future refinement. One notable simplification lies in the representation of pest dynamics. By implicitly modeling pest pressure and focusing primarily on natural enemy abundance, we may overlook the feedback loops between pest populations, natural enemy responses, and farmer management decisions. A more explicit incorporation of pest population dynamics, including factors like pest dispersal~\cite{pest_dispersal}, different growing stages~\cite{abm_pest}, and the development of pesticide resistance~\cite{pesticide_resistence}, could provide a more realistic assessment of the long-term ecological and economic consequences of different policy interventions. Furthermore, the model currently assumes a homogeneous response of natural enemies to landscape features and pesticide exposure. In reality, the arthropod community is diverse, with different taxa exhibiting varying sensitivities and functional roles. Future iterations could benefit from a more species-specific or functional group-based approach to modeling natural enemy abundance and their impact on pest control services.

From an economic perspective, the current model operates on a relatively static landscape structure, with changes in semi-natural habitat occurring as exogenous discrete shifts. A more comprehensive framework could integrate farmer decision-making processes, including risk aversion~\cite{farm_size_risk_aversion}, access to information~\cite{ecological_intensification_problem}, labor constraints~\cite{labour_agriculture}, and the adoption of other sustainable farming practices~\cite{farm_size_sustainability}. Indeed, these elements could lead to the emergence of unexpected effects from an economic and ecological point of view when policy interventions are implemented, as recently studied by two of us in~\cite{moretti2024mitigating}.  Moreover, the economic analysis relies on aggregated data from the FADN, which mask some heterogeneity in farm characteristics and market conditions across different regions and farm types within the EU-27~\cite{FADN}. Incorporating regional specificity and exploring the distributional effects of different policies on various farmer groups would enhance the policy relevance of the model.  Addressing these limitations in future research would contribute to a more robust and policy-relevant understanding of the complex relationship between agricultural management practices, landscape structure, and both ecological and economic outcomes.

Finally, it is important to acknowledge that our study did not take into account the potential negative impacts of broader ecosystem degradation driven by external factors such as climate change~\cite{pest_damage_climatechange} and invasive species~\cite{invasive_species}. These factors can only reinforce the need to adopt ecological restoration as a crucial adaptation strategy for  all farm sizes, reinforcing the urgency for policymakers to invest in research and provide solid support for such transitions.

\subsection*{Conclusion}

In conclusion, our analysis highlights the critical role of both pesticide management and landscape structure in shaping ecological and economic outcomes in agricultural systems, with farm size emerging as a key determinant of the effectiveness of ecological intensification strategies. While reducing pesticide use generally benefits natural enemy populations, the optimal configuration of semi-natural habitats for maximizing these benefits and ensuring economic viability varies significantly between small and large farms. This suggests that a nuanced, scale-dependent approach is necessary for designing effective agri-environmental policies that aim to enhance biodiversity and promote sustainable agriculture across the diverse European farming landscape.

\subsection*{Acknowledgments}

We are grateful to Jean-Philippe Bouchaud, Andrea Roventini, Lilit Popoyan, Lucas Selva and Elisa Lorenzetti for fruitful discussions. This research was conducted within the Econophysics \& Complex Systems Research Chair, under the aegis of the Fondation du Risque, the Fondation de l’Ecole polytechnique, the Ecole polytechnique and Capital Fund Management. Michel Loreau was supported by the TULIP Laboratory of Excellence (ANR-10-LABX-41).\\

\newpage

\bibliographystyle{unsrt}
\bibliography{bibl}

\clearpage

\section{Supplementary Materials}

\subsection{Landscape generation}

Here we describe the algorithm to generate the field pattern and the semi-natural habitat (SNH) distribution over a landscape. The landscape is represented by a two-dimensional lattice of $200 \times 200$ cells, where each individual cell represents a $10\,\mathrm{m} \times 10\,\mathrm{m}$ region and can be in one of two states: cropland or SNH. The generation depends on three parameters: the average field size $S$, the share of hedgerows $h$, and the share of grassland $g$.

\begin{enumerate}
    \item The algorithm starts from a landscape consisting only of cropland. The first step involves the creation of the field pattern distribution using a discrete Voronoi algorithm. From the average field size $S$, we compute the average number of fields $n$ in our landscape as:
    \begin{equation}
        n = \left\lfloor \frac{400 \mathrm{ha}}{S} \right\rfloor.
    \end{equation}
    We then generate $n$ random points within the lattice, which serve as the initial points for the discrete Voronoi tessellation. Specifically, we create a grid for all cells in the landscape and assign each cell to the nearest initial point. Note that for smaller values of $n$, the variability between different random landscape realizations will be higher compared to larger values of $n$.

    \item Once the field pattern is established, we identify the edge margins between adjacent fields and mark them distinctly in the landscape. Specifically, a single field margin is defined as the ensemble of all edges between two neighboring fields. This is crucial for generating the hedgerow patterns that are not too dispersed in the landscape. After labeling all margins, we iteratively select a field margin. If its area, divided by the total landscape size, is less than $h$, the entire field margin is converted to a hedgerow. Otherwise, only a contiguous fraction of its area is converted until the total area of hedgerows reaches exactly a share $h$ of the total landscape area. The share of hedgerows is calculated as the total area occupied by hedgerows divided by the total landscape area. Our minimum resolution of $10\,\mathrm{m}$ implies that the minimum hedgerow area is $0.01\,\mathrm{ha}$.

    \item Following the placement of hedgerows, we proceed with a similar iterative algorithm to place grassland. Here, the units considered for conversion are the individual fields identified in the first step.
\end{enumerate}

While our landscape creation does not depend a priori on the selection of a farm size, as the three parameters described above can be independent, studying realistic agricultural landscapes reveals some consistent characteristics across different farm sizes. Indeed, regions with smaller farm sizes typically exhibit smaller fields, a higher density of hedgerows separating them, and a larger share of permanent grassland. This is because smaller farms, with limited capital, can only afford smaller machinery that does not necessitate large fields for operation. Furthermore, their lower profit margins result in smaller opportunity costs for maintaining existing permanent grassland, and they often benefit from related subsidies. Finally, smaller farmers tend to value the aesthetic appeal and ecological sustainability of their landscape, thus preserving hedgerows.

In contrast, landscapes managed by large farms, benefiting from substantial capital, can acquire large machinery that leads to higher profits. However, larger machinery requires larger fields, often necessitating the removal of hedgerows. Moreover, the opportunity costs of maintaining grassland become too high, leading to the conversion of much of it to productive agricultural area.

These features are strongly evident in data (as shown in Fig.~\ref{fig:fitted} and Fig.~\ref{fig:calibration}) and represent a cross-country pattern. For this reason, the information about farm size alone can be sufficient to characterize all three parameters governing the landscape creation. Specifically:

\begin{equation}
    S = S_0 L^a
\end{equation}
\begin{equation}
    g = g_0 L^b
\end{equation}
\begin{equation}
    h = h_0 L^c
\end{equation}
where $L$ is farm size. Here, the parameters for the field-farm area relationship are taken from a recent empirical analysis~\cite{field_farm_relationship}. The parameters for the $g$ relationship are fitted to data as explained below. While a single estimation for the $h$ relationship is lacking in the literature, we can derive an estimate by observing the historical decrease in hedgerow numbers corresponding to increasing farm sizes~\cite{hedgerows_evolution_france}.

\subsection{Calibration procedure and parameters value} 
\label{app:calibration_parameters}

Model parameters were categorized into four distinct groups (see Table~\ref{tab:params}): (i) \textit{measured} parameters (\textcolor{measuredColor}{M}), directly fitted from relevant datasets; (ii) \textit{estimated} parameters (\textcolor{estimatedColor}{E}), derived from existing literature; (iii) \textit{calibrated} parameters (\textcolor{calibratedColor}{C}), fine-tuned through a calibration process to ensure model-dependent variables closely match observed data, and (iv) \textit{fixed} parameters (\textcolor{fixedColor}{F}), set manually to explore alternative scenarios that could not be excluded a priori.

\begin{table*}[t!]
    \centering
    \begin{tabular}{lcllc}
        \multicolumn{5}{c}{~} \\ \hline
        \toprule
        Section & Notation & Description & Value & Group \\ 
        \midrule 
        Structural &  & Landscape size \footnotesize{$[\mathrm{ha}]$} & 400 &  \\
         parameters &  & Landscape resolution (single cell area) \footnotesize{$[\mathrm{ha}]$} & 0.01 &  \\ 
         & $a$ &  Exponent of the farm size-field area relationship & 0.4 & \textcolor{measuredColor}{M} \\
         & $b$ & Exponent of the  farm size-grassland share relationship & -0.2 & \textcolor{measuredColor}{M} \\
        & $c$ & Exponent of the  farm size-hedgerows share relationship & -0.5 & \textcolor{estimatedColor}{E} \\ 
         & $S_0$ &  Scale of the farm size-field area relationship & 0.4 & \textcolor{measuredColor}{M} \\
         & $g_0$ & Scale of the farm size-grassland share relationship & 0.1 & \textcolor{measuredColor}{M} \\
        & $h_0$ & Scale of the farm size-hedgerows share relationship & 0.15 & \textcolor{estimatedColor}{E} \\
        \cdashline{1-5}
        Ecological & $r$ & Growth rate of natural enemies \footnotesize{$[ \text{day}^{-1}]$} & 0.01 & \textcolor{estimatedColor}{E} \\
        parameters & $\mathrm{sn}_\mathrm{ref}$ & Reference semi-natural habitat area for overwintering \footnotesize{$[\mathrm{ha}]$} & 0.5 & \textcolor{fixedColor}{F} \\
        & $\kappa_{\mathrm{SN}}$ & Carrying capacity coefficient in semi-natural habitats \footnotesize{$[\mathrm{ha}^{-1}]$} & 5000 & \textcolor{estimatedColor}{E} \\
        & $\kappa_{\mathrm{C}}$ & Carrying capacity coefficient in cropland \footnotesize{$[\mathrm{ha}^{-1}]$} & 10000 & \textcolor{estimatedColor}{E} \\
        & $\pi_{\mathrm{ref}}$ & Reference pesticide application rate \footnotesize{$[\text{\euro} \cdot \mathrm{ha}^{-1} \cdot \text{year}^{-1}]$} & 80 & \textcolor{fixedColor}{F} \\
        & $q$ & Selectiveness of pesticide application  & 0 & \textcolor{fixedColor}{F} \\
        & $D$ & Characteristic dispersal distance of natural enemies \footnotesize{$[\mathrm{ha} \cdot \text{year}^{-1}]$} & 0.2 & \textcolor{estimatedColor}{E} \\
        \cdashline{1-5}
        Economic & $y_{0}$ & Baseline potential yield \footnotesize{$[\mathrm{q}\cdot \text{ha}^{-1}]$} & 51 & \textcolor{calibratedColor}{C} \\
        parameters & $y_{1}$ & Potential yield coming from fertilizer use \footnotesize{$[\mathrm{q}\cdot \text{ha}^{-1}]$} & 18 & \textcolor{calibratedColor}{C} \\
        & $L_{\mathrm{ref},y}$ & Reference farm land area for yield scaling \footnotesize{$[\text{ha}]$} & 25 & \textcolor{calibratedColor}{C} \\
        & $\eta$ & Maximum yield loss due to pest & 0.3 & \textcolor{measuredColor}{M} \\
        & $\pi_{0}$ & Baseline crop protection expenditure (CPE) \footnotesize{$[\text{\euro} \cdot \text{year}^{-1} \cdot \text{ha}^{-1}]$} & 120 & \textcolor{measuredColor}{M} \\
        & $L_{\mathrm{ref},\pi}$ & Reference farm land area for CPE scaling \footnotesize{$[\text{ha}]$} & 25 & \textcolor{measuredColor}{M} \\
        & $\mathcal{C}_0$ & Operational costs (no CPE but with depreciation) \footnotesize{$[\text{\euro} \cdot \text{year}^{-1} \cdot \text{ha}^{-1}]$} & 750 & \textcolor{measuredColor}{M} \\    
        & $\mathcal{C}_g$ & Maintenance cost for permanent grassland  \footnotesize{$[\text{\euro} \cdot \text{year}^{-1} \cdot \text{ha}^{-1}]$} & 250 & \textcolor{measuredColor}{M} \\
        & $\mathcal{C}_h$ & Maintenance cost for hedgerows \footnotesize{$[\text{\euro} \cdot \text{year}^{-1} \cdot \text{ha}^{-1}]$} & 1000 & \textcolor{measuredColor}{M} \\
        & $p$ & Crop price per unit \footnotesize{$[\text{\euro}  \cdot \text{q}^{-1}]$} & 15 & \textcolor{measuredColor}{M} \\
        & $\ell_0$ & Baseline labor input per area \footnotesize{$[\mathrm{AWU} \cdot \text{ha}^{-1}]$} & 0.01 & \textcolor{measuredColor}{M} \\
        & $\ell_1$ & Additional labor input per farm \footnotesize{$[\mathrm{AWU} \cdot \text{ha}^{-1}]$} & 0.6 & \textcolor{measuredColor}{M} \\
        & $L_{\mathrm{ref}, \mathcal{L}}$ & Reference farm land area for labor scaling \footnotesize{$[\text{ha}]$} & 5 & \textcolor{measuredColor}{M} \\
        & $S_{\mathrm{ref}}$ & Reference field size area for labor scaling \footnotesize{$[\text{ha}]$} & 1 & \textcolor{measuredColor}{M} \\ 
         \hline
    \end{tabular}
    \caption{Parameters of the model. We indicate their units, value, and estimation method.}
    \label{tab:params}
\end{table*}

\begin{figure*}[t!]
\centering
\includegraphics[width=\textwidth]{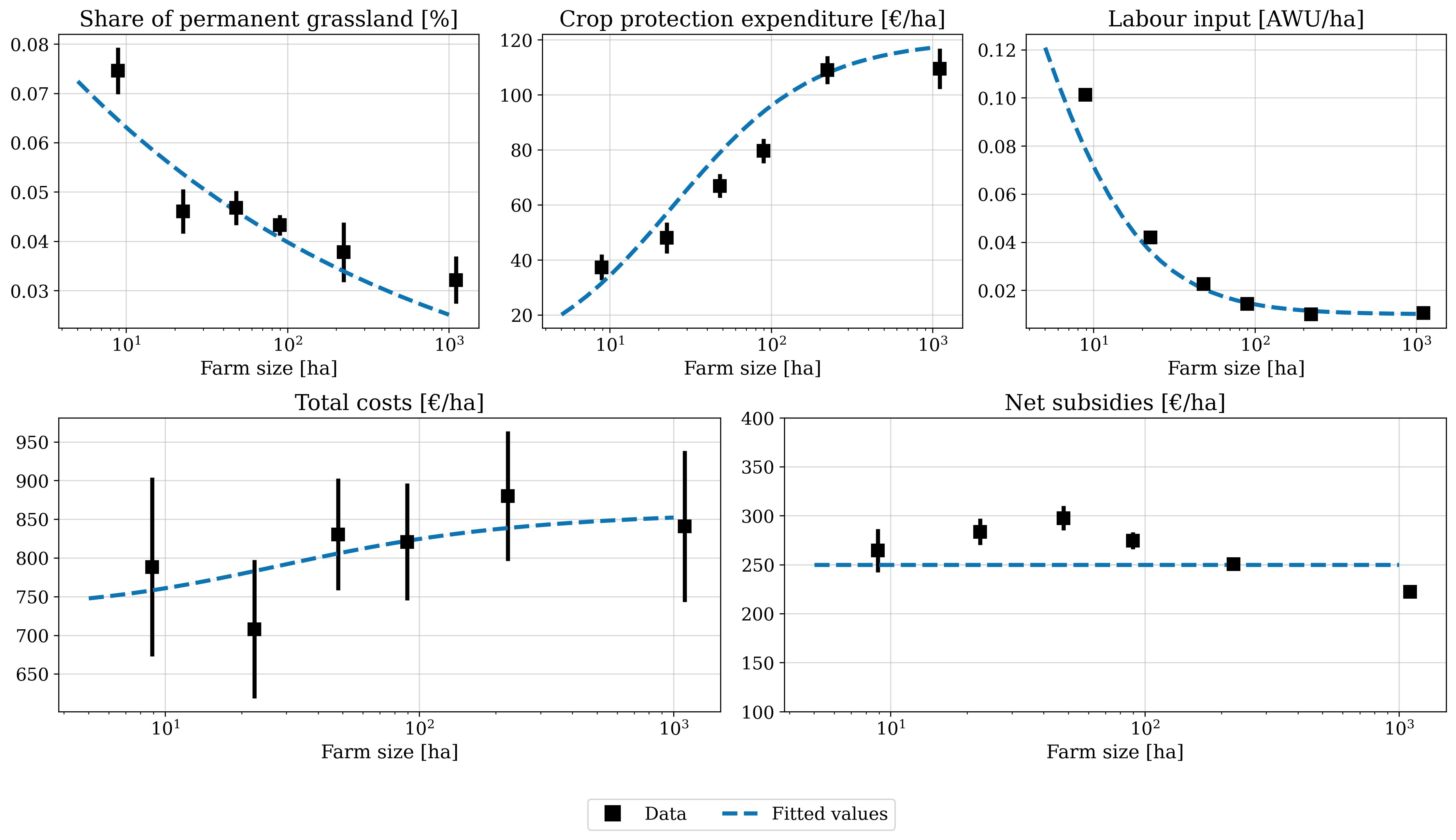}
\caption{Fitted relationships between farm size and key agricultural parameters: share of permanent grassland (\%), crop protection expenditure (which is an indicator of pesticide use intensity - 
\euro/ha), labor input (Annual Work Units per hectare - AWU/ha), total costs (\euro/ha), and net subsidies (\euro/ha). Black squares represent the weighted average of observed data from the European Farm Accountancy Data Network (FADN)~\cite{FADN} for the EU-27 region between 2018 and 2022, with error bars indicating the standard deviation of year-on-year variation. The dashed blue lines illustrate the fitted trends used as exogenous inputs for the subsequent ecological-economic modelling. These fits capture the general tendencies observed in the FADN data, such as the decreasing share of permanent grassland, increasing pesticide use and labor input with increasing farm size. Total costs show a slight positive trend with farm size, while net subsidies appear relatively stable across different farm sizes. These fitted relationships provide a realistic representation of the structural and economic characteristics of European farms of varying scales, forming the basis for simulating the impact of different policy scenarios.}
\label{fig:fitted}
\end{figure*}

\textit{Measured} parameters (\textcolor{measuredColor}{M}) are those whose influence in the model is confined to specific functions for which corresponding data are available. This category primarily includes economic parameters such as crop prices ($p$), operational costs ($\mathcal{C}_0$), subsidies ($\mathcal{S}$), and labor input ($\mathcal{L}$). These parameters were fitted using EU-27 average values for the 2018-2022 period, obtained from the European Farm Accountancy Data Network (FADN). The fitting procedure involved minimizing the least square error over a reasonable range of values. It is important to note that the objective was not to perfectly replicate the curves but rather to capture the underlying trends relevant to our study. In this regard, all fitted curves adequately represent the empirical data for each parameter (see Fig~\ref{fig:fitted}).

\textit{Estimated} parameters (\textcolor{estimatedColor}{E}) were assigned values based on empirical information and commonly reported values found in the literature. This category mainly encompasses ecological parameters. For instance, the carrying capacity of natural enemies (NE) was determined using empirical data on average individual numbers and body mass of natural enemies (NE)~\cite{carrying_capacity_parameter, carrying_capacity_parameter_2}. Similarly, information from field-based studies relevant to European cereal systems was used to inform the growth rate of NE $r$~\cite{abm_pest}. For the dispersal distance ($\gamma$), a plausible range was considered, acknowledging its variability depending on the specific taxa and species.

\begin{figure*}[t!]
\centering
\includegraphics[width=\textwidth]{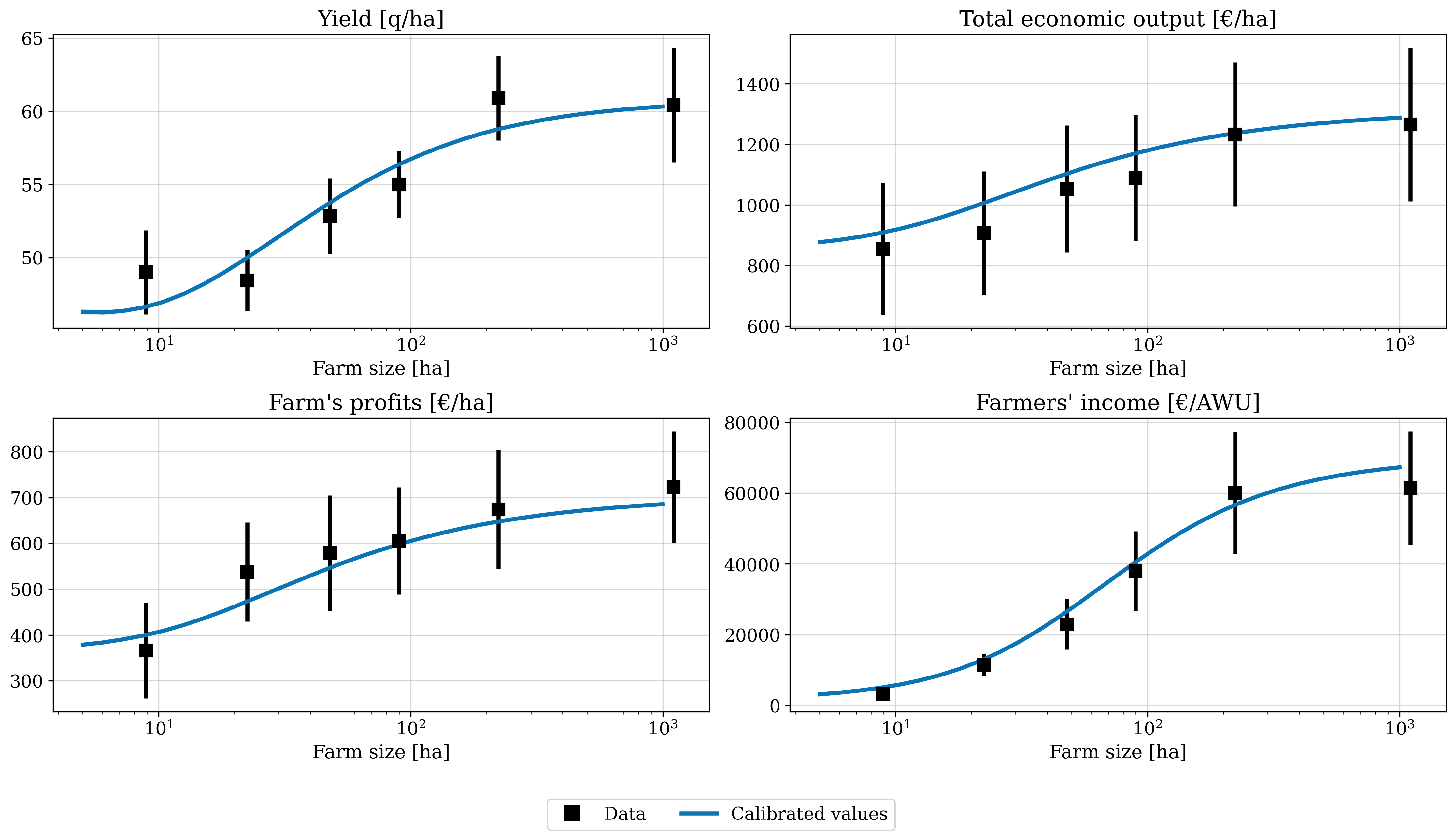}
\caption{Calibrated relationships between farm size and key agricultural output and economic parameters: Yield (q/ha), Total economic output (\euro/ha), Farm's profits (\euro/ha), and Farmers' income (\euro/AWU). Black squares represent the weighted average of observed data from the European Farm Accountancy Data Network (FADN) for the EU-27 region between 1990 and 2021, with error bars indicating the standard deviation. The solid blue lines illustrate the calibrated model outputs, achieved by adjusting \textit{calibrated} parameters (see Table~\ref{tab:params}) to best reproduce the observed trends. These calibrated relationships form a crucial part of the model validation, demonstrating its ability to represent real-world agricultural trends and providing a reliable basis for simulating the impacts of different policy scenarios on farm performance.}
\label{fig:SPLMAT_calibration}
\end{figure*}

Parameters in the \textit{calibrated} (\textcolor{calibratedColor}{C}) category are primarily associated with the yield response function, where the ecological model needs to produce realistic yield behavior. These parameters are intrinsically linked to the model's internal dynamics and therefore could not be directly fitted to external data. To align these parameters with observed data, we employed a sampling-based calibration approach. For each parameter, a realistic range of variation was hypothesized, and then a least-squares minimization procedure was performed to find the parameter values that best reproduced the observed trends. The hypothesized ranges for each calibrated parameter are detailed in Table~\ref{tab:calibration}. Given the high dimensionality of the parameter space, a direct sampling approach would have been computationally prohibitive, requiring over one million model runs for just ten values within each parameter range. To efficiently explore the parameter space, we utilized a Sobol sequence~\cite{Sobol}, which allows for a more uniform coverage and thus higher precision with fewer computational resources. We generated 4096 unique parameter sets and, for each combination, ran 100 Monte Carlo simulations. Subsequently, we computed the sum of squared differences between the simulated and observed data. Table~\ref{tab:calibration} presents the results of this calibration. The corresponding R-squared ($R^2$) and adjusted R-squared ($\bar R^2$)~\cite{R2adjusted} values demonstrate the model's effectiveness in capturing the observed trends in real-world data. For a visual comparison, Figure~\ref{fig:SPLMAT_calibration} illustrates the compatibility between the obtained model trends and the empirical data.

\begin{table*}[t!]
    \centering
    \begin{tabular}{lccc}
        \toprule
         & Description & Sampling range & Optimal \\
        \midrule
        $y_0$ & Baseline potential yield \footnotesize{$[\mathrm{q}\cdot \text{ha}^{-1}]$} & [30, 70] & 51 \\
        $y_{1}$ & Potential yield coming from fertilizer use \footnotesize{$[\mathrm{q}\cdot \text{ha}^{-1}]$} & [0, 40] & 18 \\
        $L_{\mathrm{ref}, y}$ & Reference farm land area for yield scaling \footnotesize{$[\text{ha}]$}  & [1, 100] & 25 \\
        \cdashline{1-4}
        $R^2$ & R-squared &  & 0.98 \\
        $\bar{R}^2$ & Adjusted R-squared &  & 0.97 \\
        \bottomrule
    \end{tabular}
    \caption{Results of model calibration.}
    \label{tab:calibration}
\end{table*}

Finally, parameters in the \textit{fixed} category (\textcolor{fixedColor}{F}) are those for which the literature does not provide a single, definitive value. These include reference pesticide application rate ($\pi_\mathrm{ref}$), pesticide application precision ($q$) and winter survival ($\mathrm{sn}_\mathrm{ref}$). For these parameters, we established reasonable initial values and subsequently conducted further analysis to assess their influence on the study's conclusions, acknowledging their context-dependent and complex nature.

\subsection{Scenario exploration for fixed parameters}
\label{app:scenario}

For parameters in the \textit{fixed} category (\textcolor{fixedColor}{F}) we perform further analysis to assess their influence on the study's conclusions, acknowledging their context-dependent and complex nature. We do this by primarily focusing on the ecological consequences of their variation. It is important to note that the following results are obtained after the system has been recalibrated to ensure consistency with the real data description, using the same methodology explained previously.

\subsubsection*{$\mathrm{sn}_{\mathrm{ref}}$: The role of semi-natural habitat area for overwintering}

The parameter $\mathrm{sn}_{\mathrm{ref}}$ establishes the reference amount of Semi-natural habitat (SNH) required to support the natural enemy (NE) population over winter. Increasing this value has two direct consequences: (i) overwintering becomes increasingly critical in determining the end-of-summer NE abundance, and (ii) the capacity of hedgerows to provide this overwintering service is diminished as they are fragmented within the landscape. Conversely, decreasing this value would lessen the importance of overwintering and enhance the relative contribution of hedgerows.

\begin{figure*}[t!]
\centering
\includegraphics[width=\textwidth]{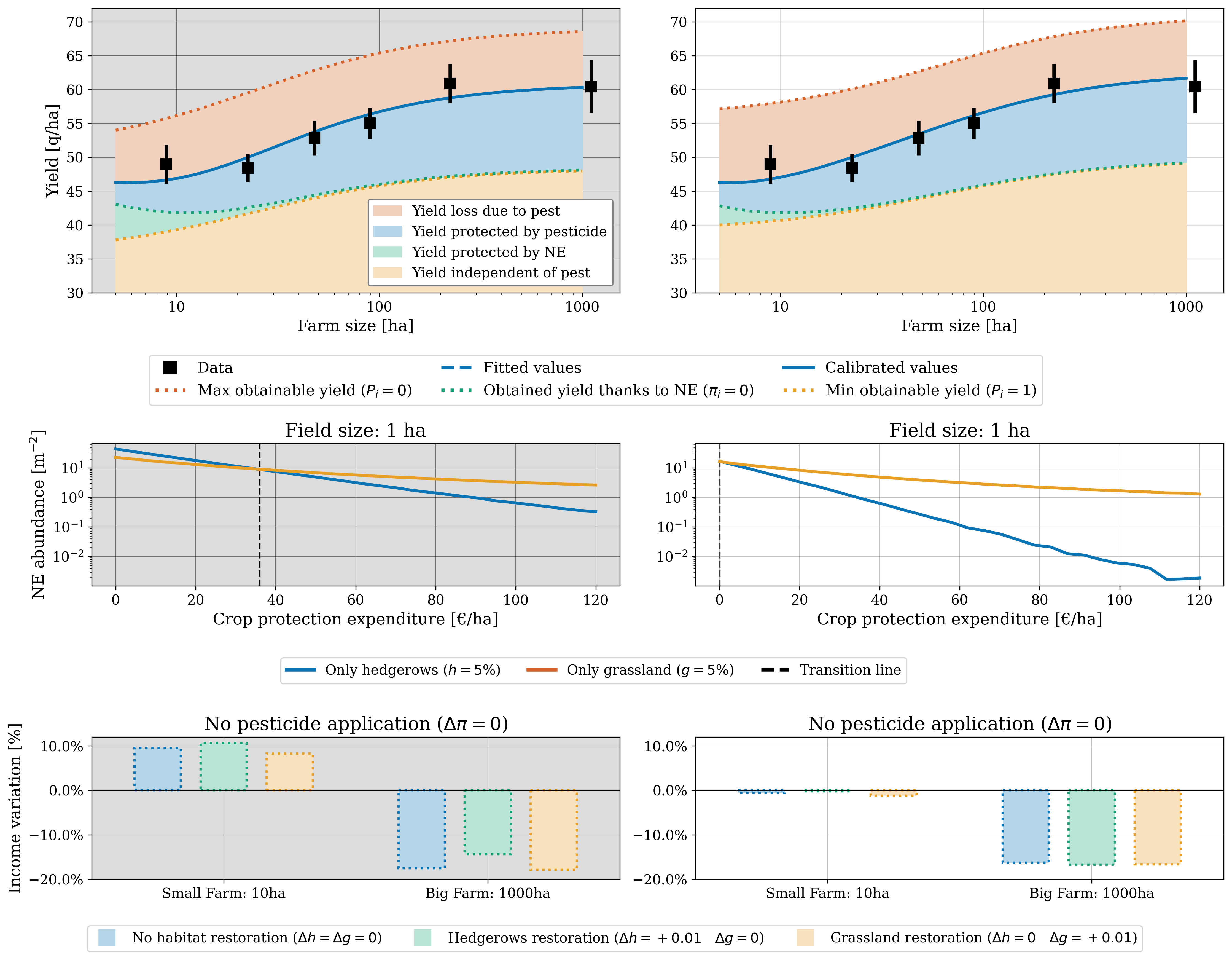}
\caption{The figure illustrates the impacts of doubling the reference semi-natural habitat area required for overwintering natural enemies from $\mathrm{sn}_\mathrm{ref} = 0.5$ ha to $\mathrm{sn}_\mathrm{ref} = 1$ ha.  The left panels display the baseline scenario results reported in the manuscript, while the right panels show the outcomes under the modified parameter value. Comparing the left and right columns reveals that a larger reference SNH area generally leads to reduced natural enemy protection and increased overall yield loss compared to the baseline scenario. Under this modified $\mathrm{sn}_\mathrm{ref}$ value, increasing grassland cover consistently proves more beneficial for mitigating these negative effects, as hedgerows alone no longer offer adequate shelter for the NE population. Consequently, policies focused solely on hedgerow restoration are less effective at improving farmers' income when pesticide use is reduced.}
\label{fig:SPLMAT_scenario_sn_ref}
\end{figure*}

To observe the impact on our study's conclusions, let us consider Figure~\ref{fig:SPLMAT_scenario_sn_ref}, where the baseline results (those presented in the main manuscript and depicted in the gray area of the figure) are compared to a scenario where $\mathrm{sn}_\mathrm{ref}$ is doubled, thus $\mathrm{sn}_\mathrm{ref} = 1\,\mathrm{ha}$. In this case, the natural enemy protection is less effective compared to the baseline scenario, and the total yield loss due to pests is higher. Regarding ecological variables, increasing grassland cover consistently proves beneficial, as hedgerows alone no longer provide sufficient shelter for the NE population. Consequently, the hedgerow restoration policy does not effectively increase farmers' income when pesticide use is reduced under this altered $\mathrm{sn}_\mathrm{ref}$ value.

\subsubsection*{$\pi_{\mathrm{ref}}$: The importance of baseline pesticide efficacy in pest control}

The parameter $\pi_{\mathrm{ref}}$ defines the baseline efficacy of pesticides in reducing the populations of both natural enemies (NE) and pests, assuming an equal impact on both. Doubling $\pi_{\mathrm{ref}}$ would thus imply a twofold decrease in pesticide effectiveness in killing both groups. Conversely, halving $\pi_{\mathrm{ref}}$ would increase the efficacy of pesticide applications for pest protection.

\begin{figure*}[t!]
\centering
\includegraphics[width=\textwidth]{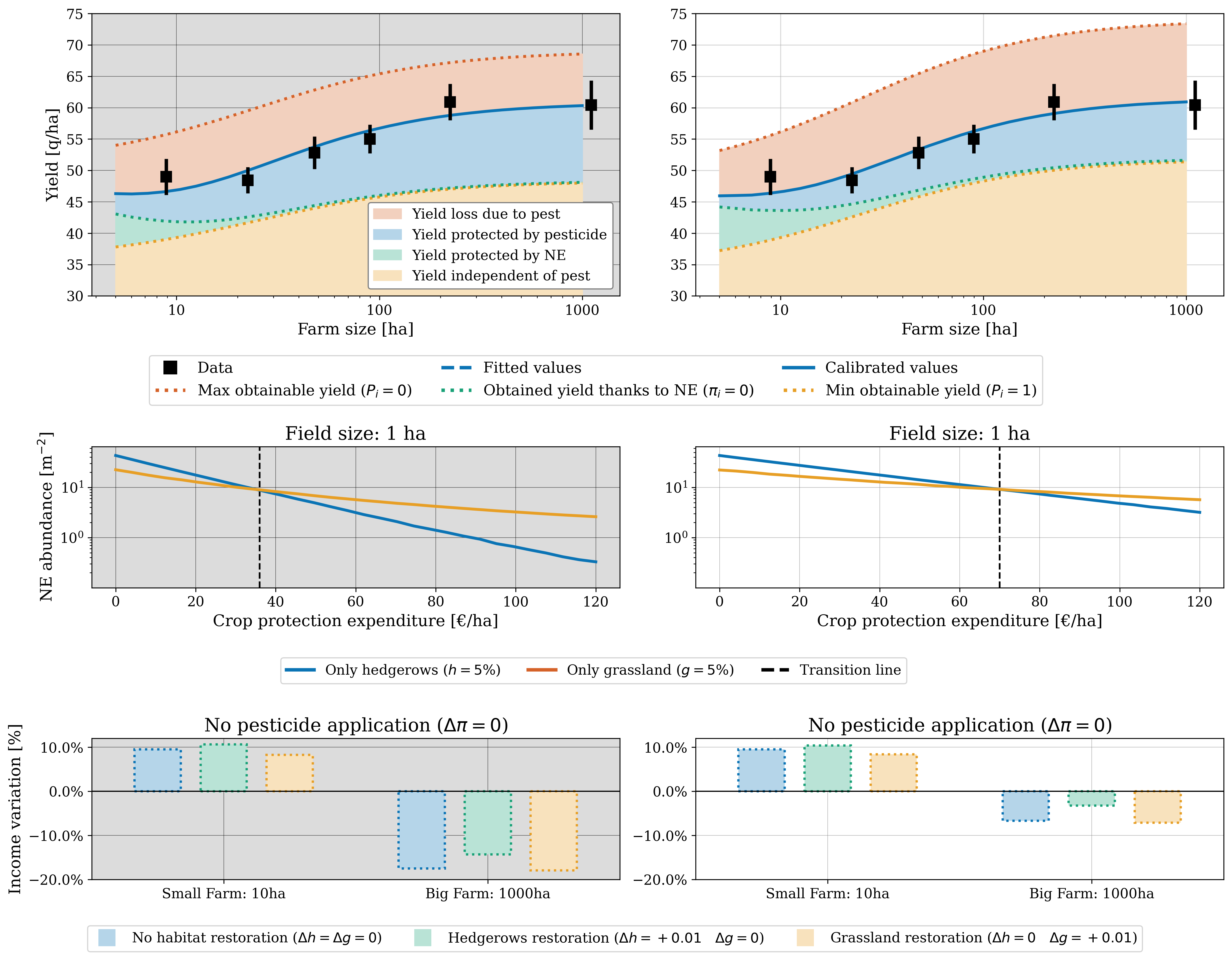}
\caption{Ecological and economic consequences of doubling the baseline pesticide efficacy from $\pi_\mathrm{ref} = 80 \; \text{\euro} \cdot \text{year}^{-1} \cdot \text{ha}^{-1}$ (left panels) to $\pi_\mathrm{ref} = 160 \; \text{\euro} \cdot \text{year}^{-1} \cdot \text{ha}^{-1}$ (right panels). The comparison shows that a higher reference pesticide efficacy generally leads to a lower reliance on pesticides for yield protection. In this modified case, the reduced pesticide efficiency results in a larger natural enemy (NE) population. Consequently, providing shelter becomes more important, making hedgerow increases a preferred strategy. Finally, transitioning to a zero-pesticide regime (right column, bottom panel) results in less significant income losses for large farms compared to the baseline $\pi_\mathrm{ref}$ (left column, bottom panel) due to the enhanced restoration potential of natural pest control.}
\label{fig:SPLMAT_scenario_pi_ref}
\end{figure*}

Considering a scenario where $\pi_{\mathrm{ref}}$ is doubled, leading to less effective pesticides, results in a greater reliance on natural enemies and landscape features for pest control (see Fig.~\ref{fig:SPLMAT_scenario_pi_ref}). Under such conditions, the advantages of enhancing natural enemy populations through semi-natural habitats, such as hedgerows and grassland, become more evident. Specifically, the transition threshold between landscapes favoring hedgerows versus grassland shifts towards higher levels of pesticide use. This indicates that hedgerows become even more crucial for sustaining higher NE population levels during winter when pesticide efficacy is reduced. Consequently, the adoption of more ecologically friendly farming practices becomes more economically favorable for farmers, underscoring the economic value of biodiversity-mediated pest control in a context of less effective chemical interventions.

\subsubsection*{$q$: The differential toxicity of pesticides to pests versus natural enemies}

The parameter $q$ quantifies the relative toxicity of pesticides to pests compared to natural enemies. A higher $q$ signifies greater selectivity, where pesticides exert a stronger negative impact on pests than on their natural predators. Conversely, a lower $q$ indicates that pesticides are relatively more harmful to natural enemies.

When the selectivity of pesticides ($q$) is increased, implying a greater reduction in pest populations relative to natural enemies, the significance of restoring and maintaining hedgerows is amplified. Hedgerows provide essential refuge and resources for natural enemies, enabling them to persist and contribute to pest control. Therefore, even if technological advancements lead to the development of pesticides with higher selectivity, the restoration of SNH and of hedgerows specifically remains a vital strategy for bolstering ecological resilience.

\begin{figure*}[t!]
\centering
\includegraphics[width=\textwidth]{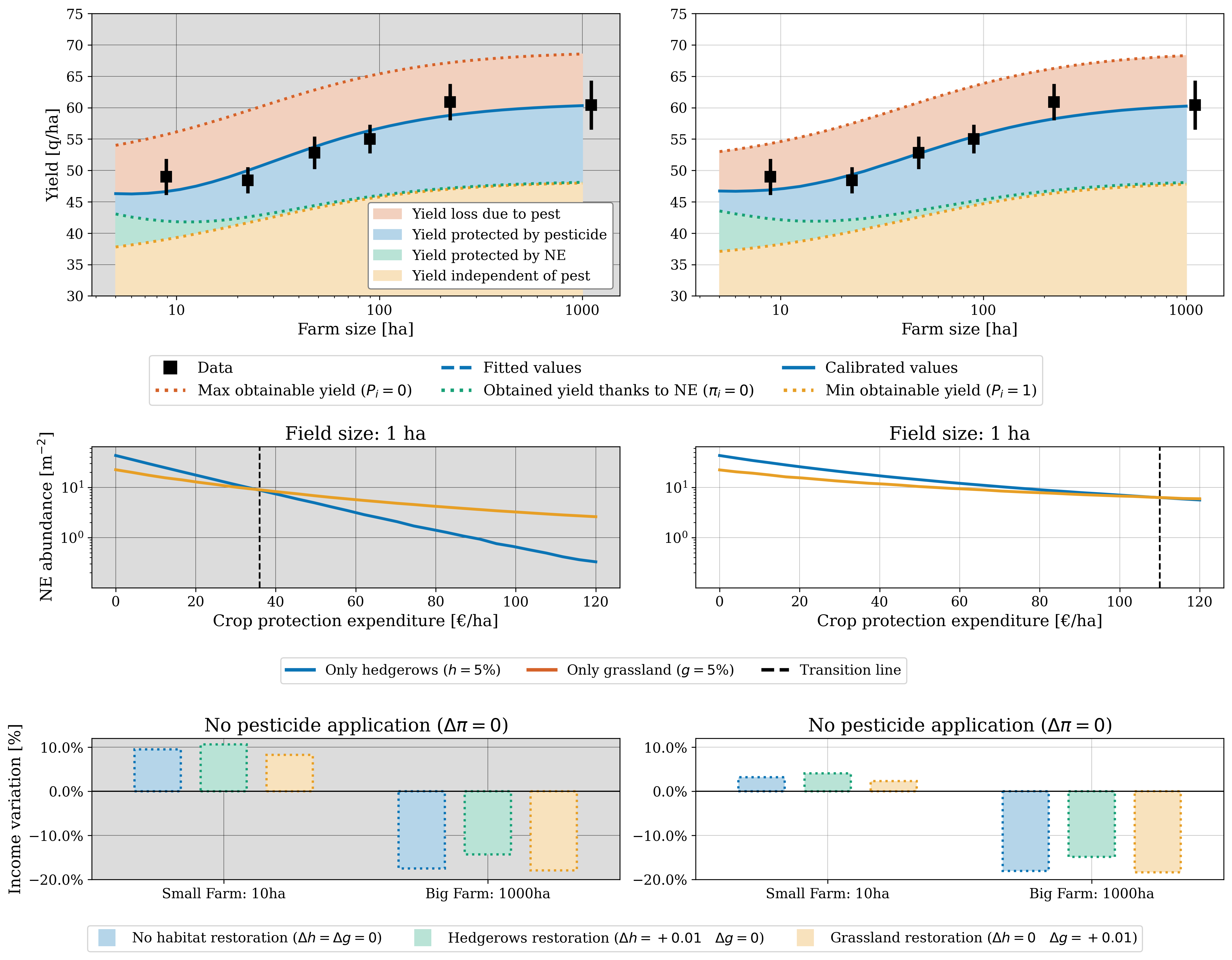}
\caption{This figure illustrates the ecological and economic consequences of choosing the pesticide selectivity $q$ equal to 0.5. The left panels display the baseline scenario results reported in the manuscript, while the right panels show the outcomes under the modified parameter value. Comparing the left and right columns reveals that increased pesticide selectivity enhances the relative importance of hedgerows in supporting NE populations and contributing to yield protection. Under a zero-pesticide regime (bottom row), the benefits of hedgerow restoration for mitigating income losses, particularly for larger farms, become are still very pronounced with higher pesticide selectivity, underscoring the complementary role of selective pesticides and SNH in sustainable pest management.}
\label{fig:SPLMAT_scenario_q}
\end{figure*}

\end{document}